\documentclass[]{spie}  %>>> use for US letter paper
\usepackage{amsmath,amsfonts,amssymb}
\usepackage{graphicx}
\usepackage[colorlinks=true, allcolors=blue]{hyperref}

\title{PANDORA: New avenues in sub-percent photometric
precision for ground based astronomy}

\author[a]{Kane M. Sjoberg}
\author[a]{Johnny Esteves}
\author[a]{Christopher W. Stubbs}
\affil[a]{Department of Physics, Harvard University, 17 Oxford St, Cambridge, MA, USA}

\authorinfo{Further author information: (Send correspondence to K.M.Sjoberg)\\K.M.Sjoberg: E-mail: sjoberg.kane@gmail.com, Telephone: 1 617 485 4360\\ }

\begin{document} 
\maketitle

\begin{abstract}
Precision photometric calibration is key to a number of astrophysical research areas such as supernova cosmology and dark energy studies. In the age of large surveys, pushing the limits of photometry is a worthy challenge, with traditional celestial calibrators limited to approximately percent precision. Here we present PANDORA, a novel calibration source that achieves sub-percent photometric precision across the optical spectrum with a selectable dynamic range and high optical efficiency. Building on proven collimated beam projector designs for relative throughput calibration of survey telescopes, PANDORA is capable of delivering photons anywhere in the 350-1100 nm spectral range. It uses selectable neutral density filters and sequential NIST-calibrated photodiode feedback to internally self-calibrate prior to operation, and is illuminated by an Energetiq EQ-99X-FC broadband laser-driven light source. We present here; an overview of the system design, operation, and laboratory characterization results. PANDORA will begin science operations this year to support the upcoming LS4 survey on the ESO 1m Schmidt telescope.

\end{abstract}

% Include a list of keywords after the abstract 

\section{INTRODUCTION}
\label{sec:intro}  % \label{} allows reference to this section

\subsection{The Importance of Precision Photometry}

High-precision photometric calibration of ground-based optical telescopes is fundamental to modern astrophysics. Virtually every area of astronomical research benefits from well-calibrated multiband photometric data, from studies of stellar populations to redshift determination of galaxies and the mapping of large-scale cosmic structure \cite{Stubbs2010}. 
A notable example of the importance of precise determination of relative throughput in multiband telescopes is the use of Type Ia supernovae to constrain the history of cosmic expansion, for which the 2011 Nobel Prize in Physics was awarded to Riess, Schmidt, and Perlmutter \cite{Riess1998}. 
% Determining the redshift of supernovae via their color requires precise relative flux measurements through two or more filters. 
Determining the luminosity distance to supernovae over a range of redshifts requires precise relative flux measurements through two or more filters. 
Uncertainty in photometric calibration is in fact the dominant source of uncertainty in error budgets constraining the equation of state parameter $w = P / \rho $ for dark energy \cite{Regnault2009}. Moreover, determining variation with redshift in the equation of state parameter $w$ will require measurements of supernova flux at sub-percent precision \cite{Stubbs2010}.

In an age of large surveys, the issue of reducing instrumental error budgets is becoming increasingly urgent. The existing SNe Ia catalog driven by current photometric surveys such as the Dark Energy Survey \cite{Brout2019} and the Subaru Hyper-Suprime Cam \cite{Toshikage2024} numbers in the hundreds; the Zwicky Transient Facility is expected to reach order $10^4$ by the conclusion of survey operations \cite{Dhawan2022}. The initiation of science operations with the Vera C. Rubin Observatory is expected to increase our catalog of Type Ia supernovae by another order of magnitude, to ~$10^5$ detections out to $z\approx 0.8$ \cite{Lochner2022}. Consequently, the capacity of high-precision photometric calibration to constrain cosmological parameters will increase as the statistical uncertainty continues to shrink below the dominant uncertainty of relative system throughput.

\subsection{Methods of Photometric Calibration}

Current photometric calibration methods can be broadly divided into two approaches: classical celestial calibration and laboratory-based metrology. The classical approach relies on observations of celestial standard sources. In this method, standard stars whose fluxes are presumed to be well known are observed at similar airmasses as the science targets to correct for atmospheric extinction and instrumental effects. Although this method has been successful in achieving a relative precision of roughly 1\%, its limitations are well documented. Variability in atmospheric conditions, uncertainties in the spectral energy distributions of the standard stars, and the inherent assumptions made in transferring these calibrations across different instruments collectively contribute to systematic errors that are challenging to reduce further \cite{Linder2009}.

In contrast, laboratory metrology offers a promising alternative. One influential approach, introduced by Stubbs and Tonry \cite{Stubbs2006}, is the transfer of quantum efficiency  measurements from NIST-traceable, precision-calibrated photodiodes to the astronomical instrument's spectral response function. This metrology transfer decomposes the calibration challenge into two components: (1) determining the instrumental sensitivity using calibrated detectors, and (2) measuring the atmospheric transmission independently. A review of the latter is beyond the scope of this work. Implementations of NIST QE-based photometric reference systems for major telescopes have demonstrated that laboratory metrology can achieve the precision required for cutting-edge cosmology, as will be discussed below. 

\subsection{A Review of Collimated Beam Projectors}
The method of photodiode-based precision photometric calibration has manifested primarily in the form of instruments called Collimated Beam Projectors (here onward referred to as `CBPs'). Broadly speaking, CBPs are variations of an instrument typically composed of a tunable monochromatic light source, direct photodiode-based monitoring along the optical path, and collimation optics that project a monochromatic collimated beam of photons. Through self-calibration routines, the throughput of a CBP can be determined to high precision, such that the dose of outgoing photons during operation is known to high precision. By comparing the photons received at the focal plane to the dose computed from simultaneous internal monitoring measurements by the NIST photodiodes, and iterating over wavelength, a precise system throughput function can be attained (see e.g., Refs.~\citenum{Souverin2024, Coughlin2018}).

CBPs provide a number of advantages compared to celestial calibration methods as a result of their self-contained design. They can be applied in-dome, projecting beams into the telescope pupil on demand. They enable a more systematic and repeatable calibration process that can be applied throughout the duration of an observatory's science operations, ensuring the calibration of long-term temporal variations in the spectral response function of the telescope to high precision. Additionally, they are efficient in that these calibrations can be performed at any time, reducing allocations of prime observing time to commissioning and calibration observations. 

\begin{figure}[ht]
  \centering
  \includegraphics[width=0.8\textwidth]{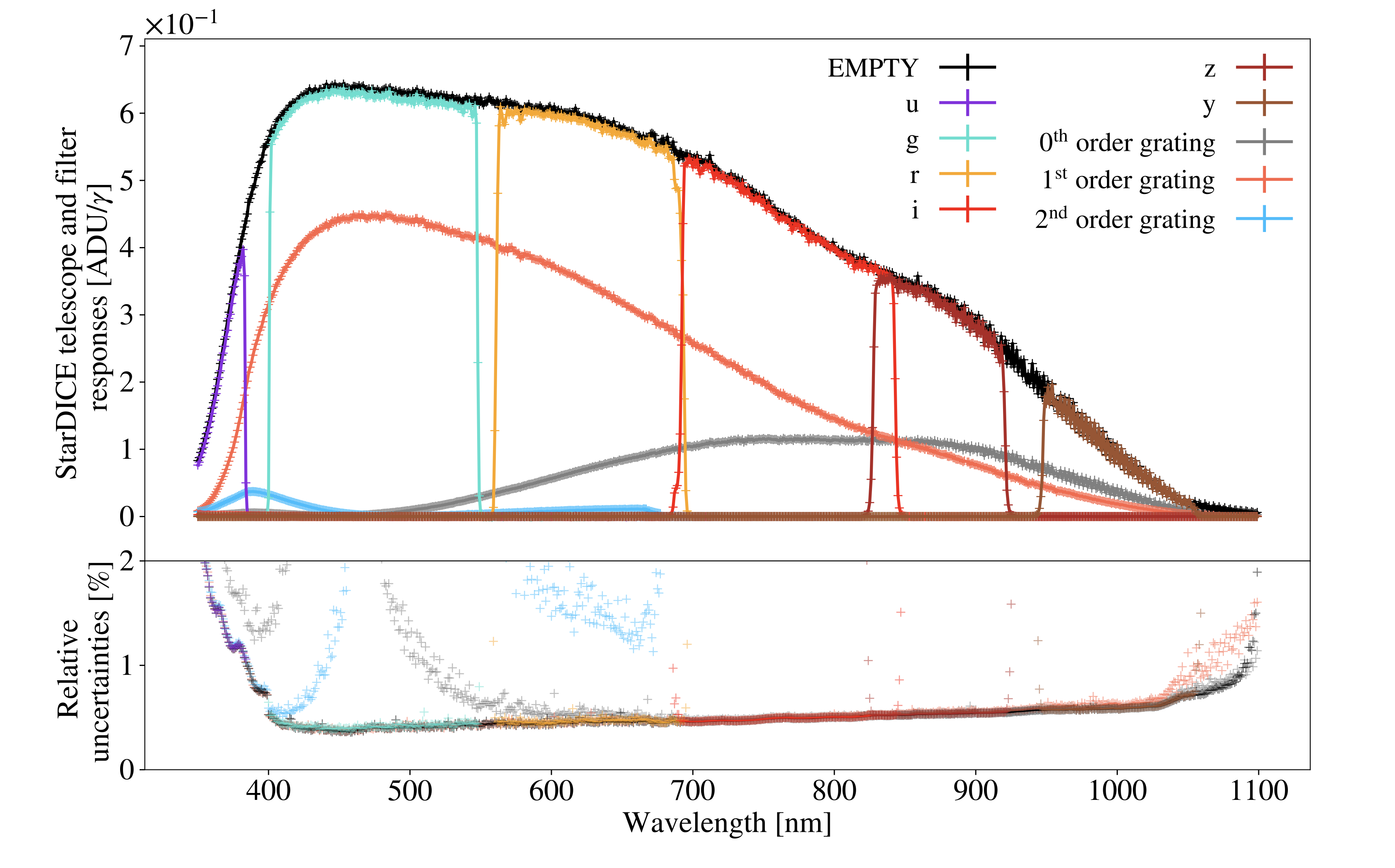}
  \caption{
    \textbf{Top:} StarDICE response against wavelength in nanometer, with the empty filter wheel slot; all \textit{ugrizy} filters; and the 0th, 1st and 2nd order diffraction of the grating. All of these responses have been measured with the 75~$\mu$m pinhole. 
    \textbf{Bottom:} Relative uncertainties over the StarDICE response measurements against wavelength in nanometer.
  }
  \label{fig:stardice_throughput}
\end{figure}

The most recent collimated beam projector designs have achieved remarkable results in attaining sub-percent precision in determining relative throughput functions of survey telescopes. Ref~[\citenum{Souverin2024}] were able to measure \textit{ugrizy} filter transmission in the StarDICE telescope to an uncertainty of $0.5\%$ per 1nm bin, as well as detecting out-of-band leakages at the $10^{-4}$ relative level. Figure~\ref{fig:stardice_throughput} shows a plot from the same paper that details the final relative throughput measurements of the StarDICE filter system.

\subsection{Limitations of Current CBP Designs}

While successful, current CBPs suffer from a number of limitations. First, they are optically inefficient. CBPs such as the system employed by Ref.~[\citenum{Souverin2024}] to calibrate StarDICE employ integrating spheres to scramble polarization effects necessary for calibrating astronomical interference filters, which are sensitive to linear polarization. However, these integrating spheres apply a geometric attenuation factor proportional to the interior surface area. Additionally, the spheres are the coupling site to the NIST photodiode. As a result, a minimum interior radius on the order of inches is imposed to ensure sufficient isotropic scattering. Coupled with the exit of the integrating sphere being a prime site for the implementation of pinhole masks to create the PSFs that are reimaged by the main telescope, the result is a geometrically defined attenuation factor in throughput that can be as high as factor of $10^6$ loss. Consequently, extremely high-power `death ray' lasers are required to drive the system: Ref.~[\citenum{Souverin2024}] use an Ekspla NT252 tunable laser that delivers 1mJ in a single $\sim$ns pulse, capable of being driven at 1kHz. These lasers are both delicate instruments to operate, often require precise thermal stabilization, and are extremely expensive and large. However, even with such powerful lasers, the calibration current produced by the NIST photodiode is in the nA to pA ($10^{-12}$ Amps) range, a level that can present an SNR challenge even when reading out on low-noise Keysight electrometers.

\section{Methods}

\subsection{A Next-Generation Collimated Beam Projector}
\begin{figure}[ht]
  \centering
  \includegraphics[width=0.5\textwidth]{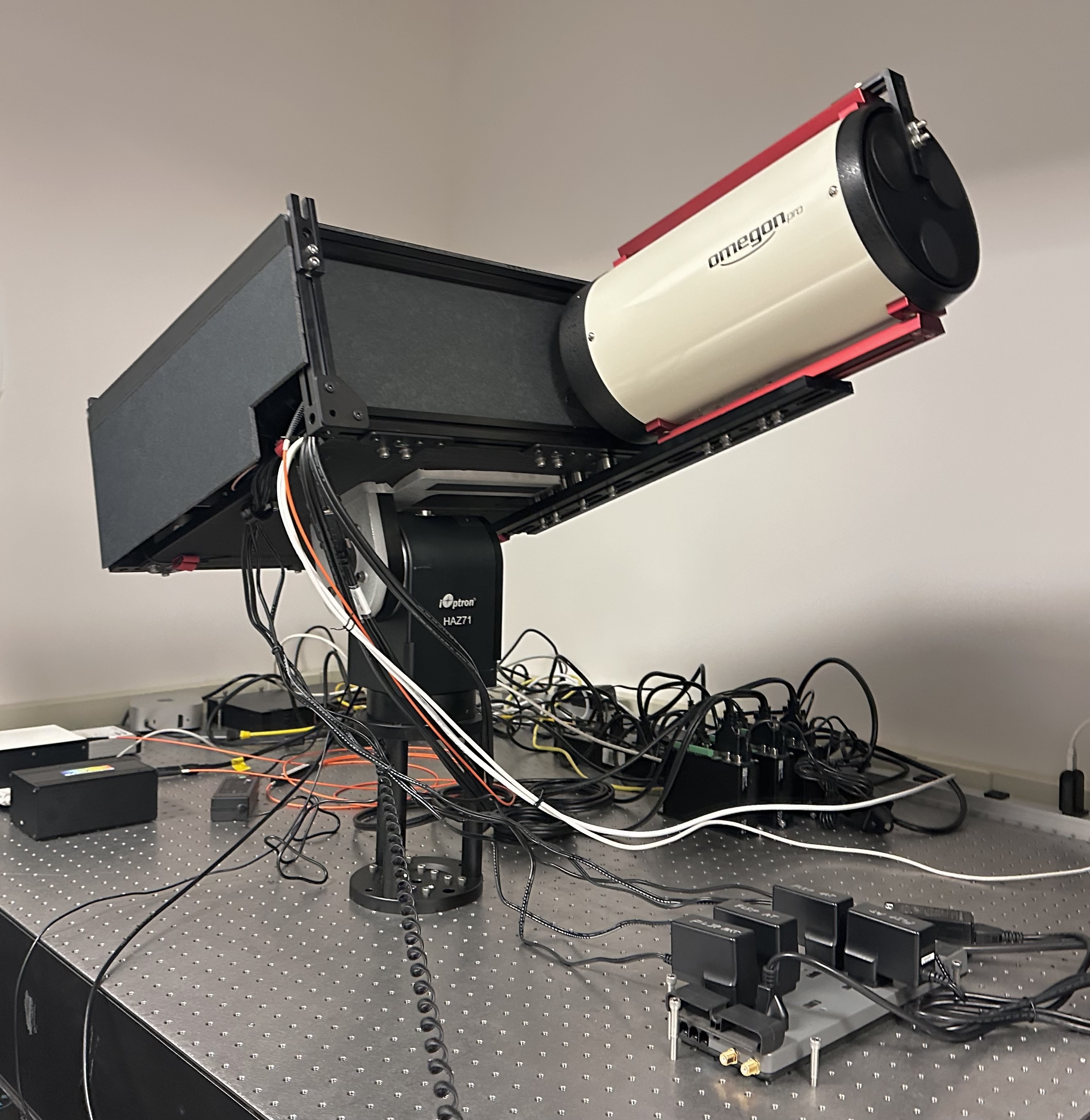}
  \caption{PANDORA on the lab bench undergoing final checks, July 2025.}
  \label{fig:pandora_on_bench}
\end{figure}

To address the aforementioned challenges and improve upon the proven capabilities of existing collimated beam projectors, we present PANDORA: \textbf{P}hotometrically-\textbf{A}djusted \textbf{N}eutral \textbf{D}ensity \textbf{O}ptical \textbf{R}ay \textbf{A}ttenuator, a novel collimated beam projector design to achieve relative photometric calibration precision at the sub-percent uncertainty level across the 350-1100 nm wavelength range. PANDORA introduces several experimental design improvements. An image of the system in-lab on the calibration bench is shown in Figure~\ref{fig:pandora_on_bench}. 

Notably, PANDORA dispenses with the use of an integrating sphere to attain vastly improved optical efficiency. This has the benefit of enabling the use of a smaller, cost-effective, and safe light source, and enabling effective dynamic range matching with the main observatory telescope. As mentioned previously, integrating spheres serve two primary purposes: erasing memory of polarization, and serving as a stable geometric constraint for the integration of the NIST-calibrated photodiode with the optical path. To ensure linear polarization is not present in the outgoing beam, PANDORA employs a quarter-wave plate as the final refractive interface in the system. To enable constant photodiode-based monitoring, PANDORA splits the optical beam into a high-flux monitoring arm, and an emission arm, using a simple uncoated UVFS plano-convex lens inserted into the beam at an angle.

Additionally, PANDORA employs a bank of selectable Neutral Density filters, allowing it to attenuate the output flux of the beam over six orders of magnitude (OD 0 to OD 6 in steps of OD 0.5), enabling dynamic range matching between PANDORA and the main telescope. The optical setup of this system is described in more detail in the following sections.

\subsection{Light Source \& Wavelength Selection}

PANDORA's light source is an Energetiq EQ-99X-FC LDLS (Fiber-Coupled Laser-Driven Light Source), outputting the same flux to within an order of magnitude over the 350-1100nm wavelength range. A sample spectrum from the light source is shown in Figure~\ref{fig:ldls_spectrum} below. This light source was chosen as a suitable match for PANDORA's high optical throughput due to its temporal stability, high brightness, and strong broadband output.

\begin{figure}[ht]
  \centering
  \includegraphics[width=0.6\textwidth]{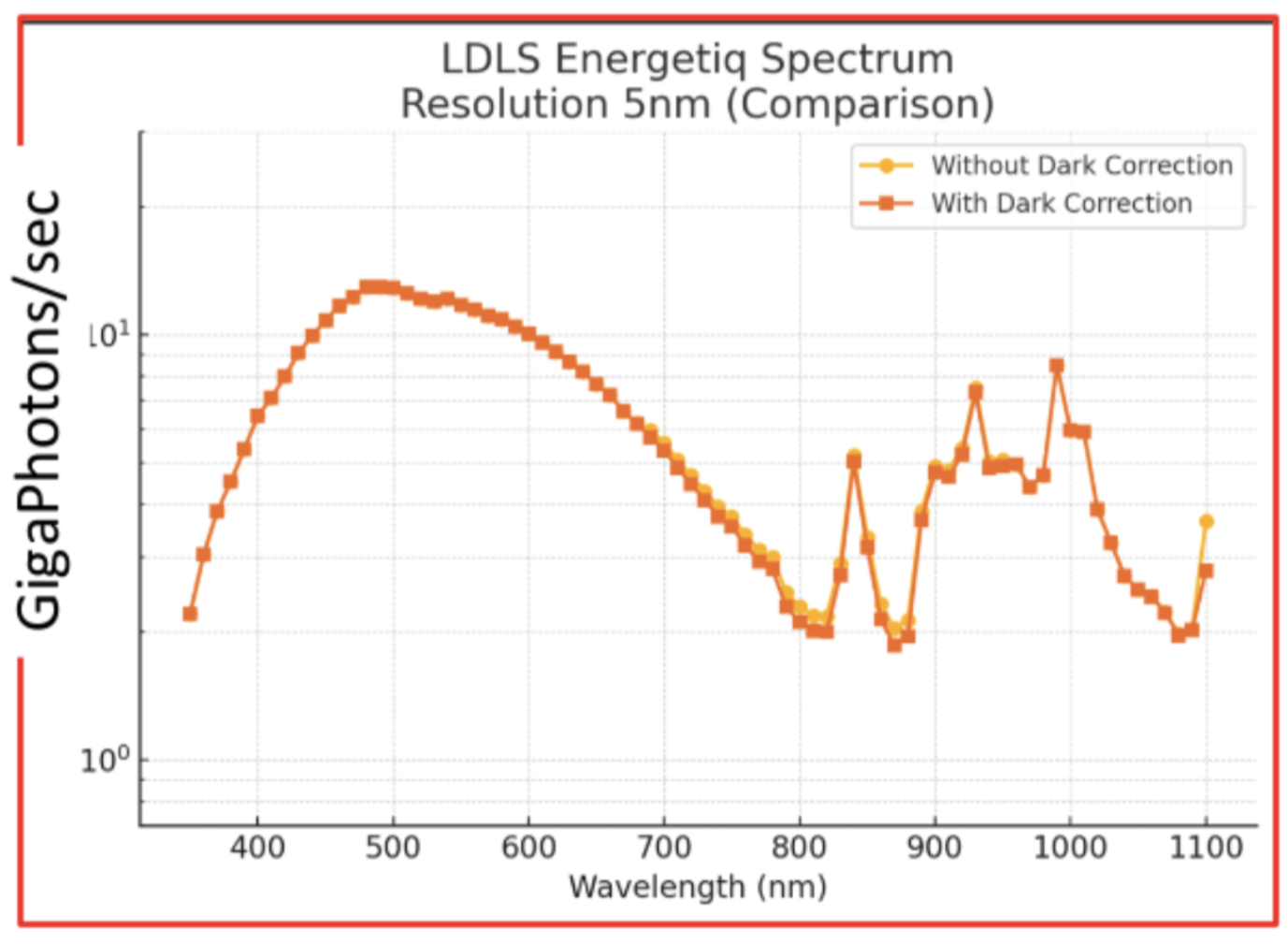}
  \caption{Energetiq EQ-99X-FC sample spectrum.}
  \label{fig:ldls_spectrum}
\end{figure}

The output of the LDLS is fed via fiber optic into a Spectral Products CM112 Double Monochromator, capable of attaining a sub-nm optical bandpass with 0.125mm slits inserted. The monochromator is controlled remotely via python scripting as well as the command line. Laboratory repeatability tests have shown the double monochromator to be capable of sub-nanometer commanded positioning repeatability. The resultant monochromatic output of the LDLS + Monochromator is passed via a $600\mu\text{m}$ mode-scrambling fiber optic cable to a Thorlabs RC12SMA-P01 Protected Silver collimator, marking the beginning of the main optical path. 

Wavelength positioning is validated via the inclusion of a StellarNet Black Comet Concave Grating Spectrograph, which possesses a maximum spectral resolution of 10 nm. The spectrograph is connected via a bifurcated optical fiber that splits the output from the monochromator between the spectrograph and PANDORA. Wavelength calibration of the spectrograph is enabled via an Ocean Optics HG-2 Mercury lamp, which feeds the spectrograph concurrently via an additional input to an additional bifurcated optical fiber.

\subsection{Beam Control \& Calibration}

\begin{figure}[ht]
  \centering
  \includegraphics[width=0.6\textwidth]{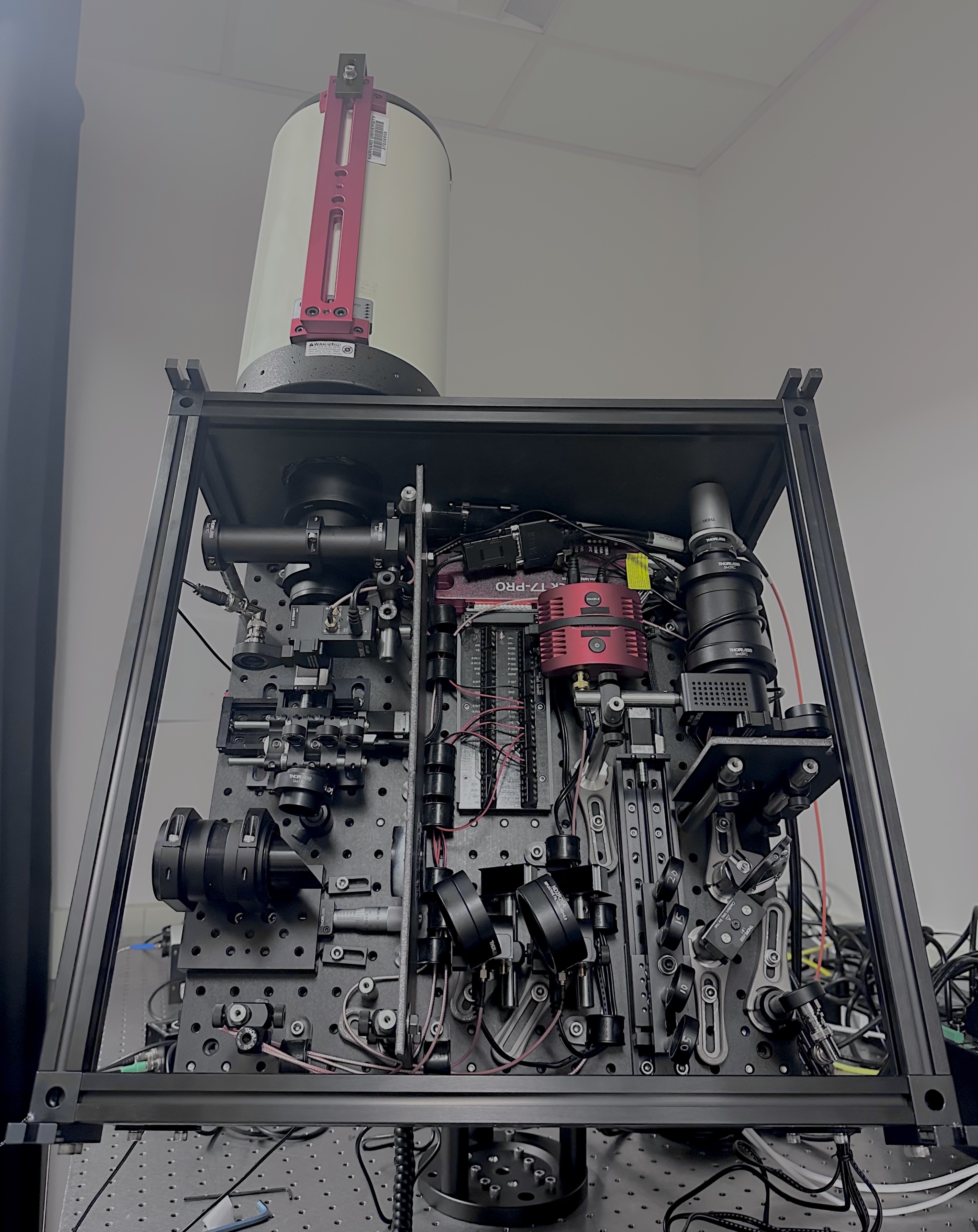}
  \caption{A view from above of PANDORA's internal optical setup. The optical path is roughly traced by a U-shape, with initial collimation occurring in the top right corner. }
  \label{fig:optical_diagram}
\end{figure}

PANDORA is distinguished from existing collimated beam projector designs by its metrology methodology and unique optical design enabling user-adjustable throughput. A view of the optical payload is shown in Figure ~\ref{fig:optical_diagram}.

The optical path is organized as follows. The light from the LDLS \& monochromator is collimated by a Thorlabs RC12SMA-P01 Protected Silver Reflective Collimator. The collimated beam is modulated by a Thorlabs SH05BT 0.5" high-speed optical shutter as well as a permanently mounted 290nm long-pass filter, as well as a removable 495nm long-pass filter mounted on a Thorlabs MFF101 flip mount. The beam is then masked by a D=10mm rounded-blade aperture to reduce stray \& uncollimated light. This beam is then split by a UV Fused Silica plano-convex lens to produce a reflected and refracted beam. The refracted beam passing through the window becomes the monitoring beam arm: $\approx 95\%$ of input flux is refracted through, where it impinges on a Hamamatsu S2281 silicon photodiode. This photodiode serves as the primary monitoring source by which a high-SNR determination of the emitted photon flux is determined. 

Meanwhile, $\approx 5\%$ of the light is reflected (following classical reflection \& refraction laws at an air-glass interface) and redirected for projection (ultimately, the delivered dose of photons). This reflected beam enters an attenuation section of the system, where a set of Thorlabs 1" reflective Nickel-coated UVFS neutral-density filters can be inserted into the beam. The filters are all optionally insertable, and are situated as follows: a set of four ND filters, from OD 0.5 - 2.0, are mounted laterally on a Zaber X-LSM150A-SE03 stage, enabling optical user-defined attenuation factors in steps of OD 0.5. Behind this stage are two additional ND 2.0 filters mounted on flip mounts. In all cases, the ND filters are angled such that light is directed into a Thorlabs LB1 beam block, ensuring stray light and secondary reflections are minimized.  Thus, the system can provide additional attenuation to match dynamic range outputs by as much as an additional factor of $10^6$ if necessary. This means that PANDORA's output flux can be modulated to match efficiently to reach high SNR on the main calibration telescope detectors in efficient exposure times. 

\begin{figure}[ht]
  \centering
  \includegraphics[width=0.9\textwidth]{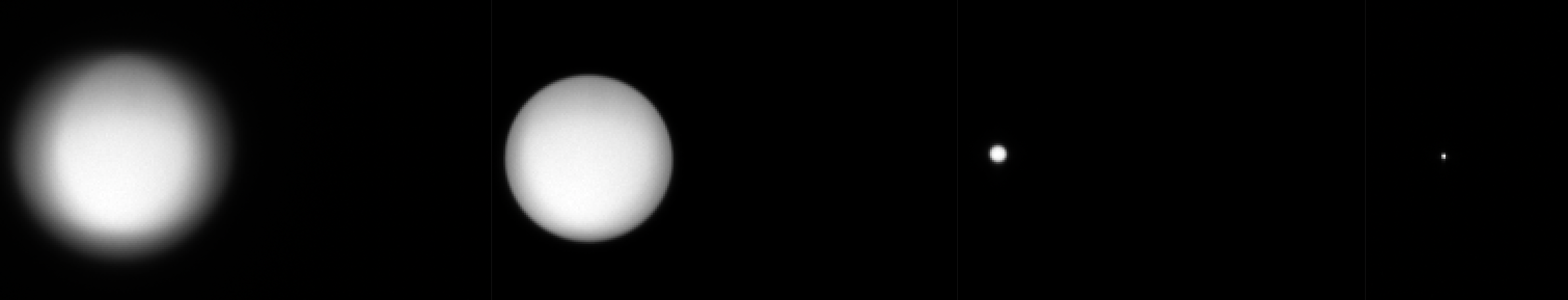}
  \caption{Shown here are images acquired at the output of PANDORA. We see in order from left to right, a view of the refocused beam without masking, then 1mm, $100\mu\text{m}$, and $10\mu\text{m}$ pinhole masks. The pixel scale of this image is $\approx 1.8 \text{ arcseconds/px}$. }
  \label{fig:pinholes}
\end{figure}

Following this attenuation section is a Newport 50328AL replicated aluminum off-axis parabolic mirror with \emph{D}=1.5 in,  \emph{FL}=4 in. The collimated beam is refocused at this point for projection into a larger collimation optic, an Omegon RC6 Ritchey-Chretien reflecting telescope with \emph{D}=154mm,  \emph{FL}=1370mm. This optic is used to project a large-diameter collimated beam such that PSF size can be effectively matched to the pixel scale of the main telescope. To further enable PSF modification and focusing of this projection telescope, a set of Thorlabs high-precision pinhole masks are placed on an xy-stage composed of Zaber X-LSM050A-SE03 (pinhole selection) and X-LSM025A-SE03 (focusing) linear stages. Thus, the collimated and attenuated beam is prepared for projection by this selectable pinhole mask design, coupling the light to the main projection optic. Shown in Figure \ref{fig:pinholes} are images of the pinhole masks positioned in-beam during PANDORA's operation. 

Located immediately before the pinhole masks is a Thorlabs SAQWP05M-700 D=10mm super-achromatic quarter-wave plate, for converting any incidental linear polarization into a fully circular state. Behind that is a second Hamamatsu S2281 photodiode which is also mounted on a Thorlabs flip mount. This second photodiode forms the basis by which comparative throughput measurements can be attained when the system acquires data in a self-calibrating mode.

\subsection{Throughput Measurements via Monitor Diodes}\label{subsec:monitor}

PANDORA is unique in that it makes use of insertable optical elements in the beam path to adjust output flux levels as necessary, and so the throughput function of the system, as well as the total propagated output flux uncertainty, is correlated to the combination of optics in the beam at any given time. This therefore requires that the wavelength-dependent throughput of every optical element past the UVFS `beamsplitter' be characterized \emph{prior} to the start of science calibration exposures. 

With two photodiodes operated simultaneously, in general, the conceptual layout for determining the throughput of optical components in the system involves: 
\begin{itemize}
   \item Setting the wavelength
   \item Measuring dark currents for both photodiodes;
   \item Recording the photocurrents, dark-calibrating the currents, and computing the throughput factor:
   \[
   \text{ThroughputFactor} = \frac{\text{Output flux (downstream photodiode)}}{\text{Monitor signal (upstream monitor photodiode)}}
   \]
   \item Performing repeat measurements at that wavelength to minimize uncertainty;
   \item Iterating over the wavelength range of interest.
 \end{itemize}

By performing this sequence of measurements with the filter of interest both in and out of the beam, one has the necessary information to determine the wavelength-dependent attenuation function of the optical element by dividing the two data vectors wavelength-wise. 

These measurements are performed prior to science operations, with the intent of building a local register of the throughput functions of every optical component in the system. Notably, PANDORA can perform a full system calibration autonomously, as the interior of the optical system is light-tight, and so a weekly or monthly self-calibration schedule could be implemented if desired.

Using this local register, PANDORA can compute an optimized exposure time and filter combination that will deliver a dose of N photons at any given wavelength. This strength of the system, effectively an enormous output dynamic range without loss of throughput precision, allows it to be unusually adaptable to a wide variety of optical systems, which can require specific photon fluxes to avoid either CCD saturation, or conversely insufficient SNR during calibration exposures. 

As discussed, PANDORA can also switch between pinhole masks of varying diameters, in order to produce PSFs that are well-matched to different focal lengths, while minimizing contamination from ghosting in the optical train of the telescope undergoing calibration. Due to the intentional lack of any refractive focusing elements in the optical path other than a planar superachromatic quarter-wave plate,  there is no wavelength dependence across different pinhole sizes, and that instead a single uniform flux measurement is sufficient to determine the flux offset between two pinhole masks. Furthermore, wavelength independence is entirely verifiable due to the presence of the aforementioned second photodiode behind the pinhole stages.

To fully characterize the wavelength-dependent throughput of PANDORA, a second calibration routine is required. A NIST-calibrated photodiode is installed at the output pupil of the masked projection optic, the Omegon RC6 projector, and throughput measurements are acquired concurrently with the high-flux photodiode. The photodiode measuring the final output of the RC6 is the only photodiode whose wavelength response contributes to the uncertainty budget of PANDORA's throughput, for in this case a referenced measurement is not being performed, but rather the wavelength dependence of the output collimated beam is being measured. It should be noted that this \emph{does not measure the total photon flux through the pupil}, as the collecting area of the  D=11.3mm calibrated Hamamatsu S2281 photodiode is only $\approx 100 \mathrm{mm}^2$, while the  D=50.8mm masked pupil of the Omegon RC6 is $\approx 20$x larger. However, full characterization of the pupil is unnecessary for wavelength-dependence measurements, because no refractive focusing elements are used in any part of the primary beam path, and thus a wavelength-dependent spatial variance in delivered flux over the Omegon RC6 pupil is not present.

With this second calibration measurement, the total combined wavelength-dependent throughput of all fixed in-beam optics is characterized, including the Omegon RC6. This calibration serves an essential purpose: determining the spectral throughput relation between the signal measured at the high-flux `monitor' photodiode located behind the plano-convex `beamsplitter' lens, and the pupil of PANDORA. Crucially, this transfer function enables knowledge of the \emph{delivered monochromatic photon dose} during science operations, simply by integrating the recorded flux that the monitor photodiode receives for the duration that PANDORA delivers a monochromatic dose of photons. This is necessary to correct for any temporal variation in our light source at the sub-percent level, resulting in a precise measurement of the total delivered photon dose.

Thus, with these two preparatory self-calibration procedures, PANDORA possesses a complete register of the throughput functions of all components in the system, enabling it to deliver monochromatic doses of photons characterized to sub-percent precision.

\subsection{Mechanical Positioning of PANDORA}
PANDORA's optical payload is entirely mounted on a heavy-duty iOptron HAZ71 Strain-Wave mount, which is computer controllable and has a load capacity of 32 kg. Consequently, the entire system is capable of being pointed at any azimuthal angle, and elevation angles as low as -40 degrees, allowing it to be easily integrated into telescope domes in a variety of positions. By virtue of these degrees of freedom, it can furthermore vary the incidence angle of the collimated beams it projects, enabling user-selectable positioning of a PSF on the focal plane of the main telescope with arcsecond accuracy.

\subsection{Software, Operation, and Data Acquisition}
Computer control of PANDORA is centralized via a single Mac Mini (M4, 2024) and a custom python class, PandoraBox, available \href{https://pandora-box.readthedocs.io/en/latest/index.html}{on GitHub}. From the command line, the states of all optomechanical components in the system can be queried, and all data acquisition is performed from the command line and stored. All of PANDORA is driven from a single shell command called \texttt{pb}. The main routine \texttt{pb measure-pandora-throughput} acquires the data shown in Section 3: it steps the monochromator in user-defined increments, interleaves dark frames, logs both Keysight electrometers, and writes a self-documenting .csv file. The .csv file contains metadata including the exposure and wavelength parameters defined by the user, as well as the dark current measurements reported by the Keysight electrometers and the throughput measurements for each exposure. Sub-commands are also available from the command line, for example
\texttt{pb set-wavelength 500},
\texttt{pb open-shutter},
\texttt{pb get-keysight-readout 2 --autoRange}, or
\texttt{pb zaber nd-filter clear} all work so that ad-hoc shell work as well as scripted automation is possible from the CLI.

Optomechanical components including the Zaber stages, Thorlabs MFF101 Flip Mounts, and the SH05BT Shutter are controlled via analog IO ports on a LabJack T7 Pro DAQ, connected to the Mac Mini via an ethernet switch. Monochromator and spectrograph control is USB-based. The HG-2 Mercury lamp is switched on and off via an ethernet-connected power switch. 

Current from the NIST-calibrated photodiodes is read out on two high-precision Keysight electrometers, with dark currents as low as $10^{-12}$ Amps. The Keysight electrometer readout parameters are set in the command line. They can be set to sample the received current at varying sampling rates, as well as cumulative samples.

\section{Results}

\subsection{Relative Filter Throughput Measurements}

\begin{figure}[ht]
  \centering
  \includegraphics[width=\textwidth]{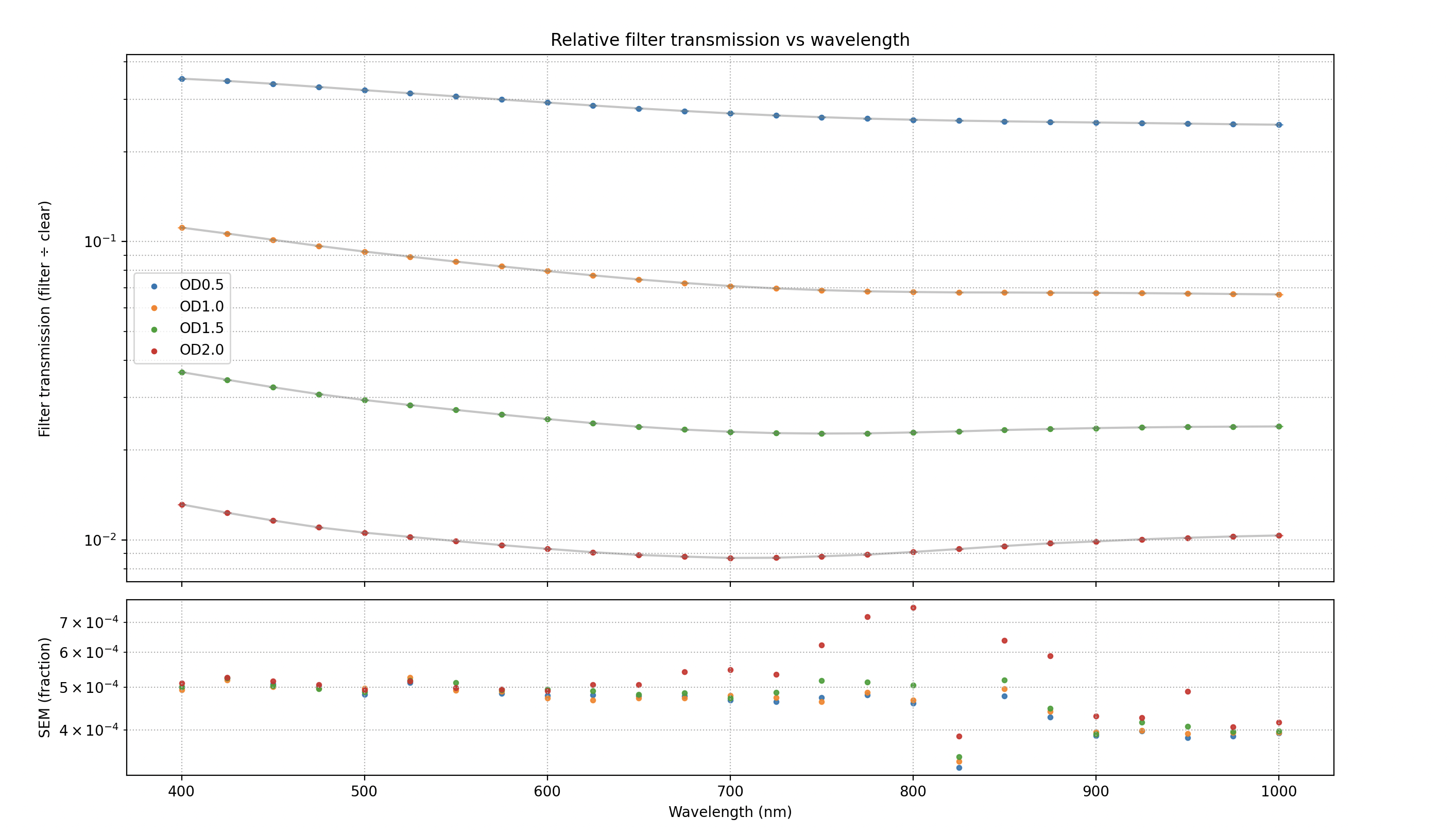}
  \caption{Shown in the top panel is the relative throughput, computed as the difference between the ratio of the monitor photodiode and the downstream photodiode. This dataset acquired five minutes of exposure time per wavelength increment, in a 1s-dark 10s-light repeating interval. The Nickel-deposited ND filters adhere well to their manufacturer-supplied attenuation factors. The bottom subpanel shows the relative uncertainty for each measurement, at each wavelength, colored by filter.}
  \label{fig:relative_throughput}
\end{figure}

Shown in Fig~\ref{fig:relative_throughput} are the relative throughput curves attained for the Neutral Density filters, via the method described in Subsection~\ref{subsec:monitor}. The Neutral Density filter transmission curves show that they adhere closely to the reported OD attenuation values.

These results, particularly the extremely low uncertainties, indicate that the system is easily capable of achieving sub-percent internal calibrations of all relevant optical components. At the time of submission, the NIST-traceable exit-pupil photodiode has not yet been delivered, and we therefore report relative transmissions and internal repeatability here, and defer absolute monitor-to-exit transfer measurements and end-to-end photon-count validation to August.

\section{Conclusion}
PANDORA has been fully integrated and is completing lab characterization tests. PANDORA is shipping to Chile in August 2025, where it will undergo commissioning before transitioning to science operations, serving as the primary photometric calibration standard for the upcoming LS4 Survey on the ESO 1m Schmidt telescope at La Silla Observatory.

\acknowledgments % equivalent to \section*{ACKNOWLEDGMENTS} 
The author would like to express their gratitude to Christopher W. Stubbs and Johnny Esteves for their time, mentorship, and support in the development of this instrument. 
We are also grateful to the US DOE (DE-SC0007881 for DOE 133341), our LS4 collaborators, Schmidt Sciences, our NIST colleagues, Energetiq, and Harvard University for their support of this project.

% References
\bibliography{bibliography} % bibliography data in bibliography.bib
\bibliographystyle{spiebib} % makes bibtex use spiebib.bst

\end{document}